\documentclass[conference,a4paper]{IEEEtran}
\usepackage{xcolor}
\usepackage{balance}

\ifCLASSINFOpdf
  \usepackage[pdftex]{graphicx}
\else
  \usepackage[dvips]{graphicx}
\fi
\ifCLASSOPTIONcompsoc
 \usepackage[caption=false,font=normalsize,labelfont=sf,textfont=sf]{subfig}
\else
 \usepackage[caption=false,font=footnotesize]{subfig}
\fi
\usepackage{adjustbox}

\usepackage[range-units = single]{siunitx}

\usepackage{booktabs}
\usepackage{mathtools}
\usepackage{array,xparse}
\usepackage{tikz}      
\usepackage[footnote]{nicematrix}
\usepackage{enumitem}

\begin{document}
%



\title{3D-Printed Dual-Polarized Magneto-Electric Dipole Antenna with Wideband High Isolation for Full-Duplex Applications}

\author{\IEEEauthorblockN{
Mehmet Ahad Yurtoglu,   
Ramez Askar   
}                                     
\IEEEauthorblockA{
Dept. of Wireless Communications and Networks,\\ 
Fraunhofer Institute for Telecommunications, Heinrich Hertz Institute, HHI, Berlin, Germany}
\IEEEauthorblockA{ \emph{\{mehmet.ahad.yurtoglu, ramez.askar\}@hhi.fraunhofer.de}}
}



\maketitle

\begin{abstract}
The paper introduces a novel dual-port dual-polarized magneto-electric dipole (MED) antenna with orthogonal Gamma and inverted-Gamma shape probes, which was fabricated by means of an additive 3D metal printing process. Electromagnetic wave simulation and RF measurement report a resonance bandwidth from 3 GHz to 4 GHz at both MED's ports with respect to a standing wave ratio of less than 2. The cross-polarization isolation (XPI) between the MED's ports was also measured to be greater than 50 dB across its entire resonance bandwidth. The paper also thoroughly examines the impact of misalignments in the polarization of the MED probes on the XPI level. The broadband resonance and excellent isolation between the MED ports make it a strong candidate for a full-duplex wireless transceiver in network infrastructure.

\end{abstract}

\vskip0.5\baselineskip

\section{Introduction}
\label{sec:introduction}
Antenna passive cancellation (isolation) is an effective technique for self-interference cancellation (SIC) in a full-duplex wireless transceiver\footnote{Note that over a decade of research evidently proved that self-interference cancellation is the realization methodology for a full-duplex duplexing wireless communication scheme.}. On the one hand, antenna SIC is principally a desirable SIC technique because it cancels the self-interference in the radio frequency (RF) domain, and it is a passive technique, i.e., it does not waste any energy for SIC purposes. On the other hand, many antenna-based SIC techniques fall short in terms of SIC amount and bandwidth, making them insufficient to rely solely on for RF SIC purposes \cite{singleChannelFD}. Therefore, antennas with a broadband isolation (passive SIC) feature are very suitable for full-duplex wireless transceivers. Additionally, the full-duplex scheme intrinsically enables a monostatic sensing feature that can operate in parallel to a bidirectional communication link -- known as integrated communication and sensing (ICAS). 

Speaking of broadband isolation, polarization mutual orthogonality in a multi-port antenna can be utilized for antenna SIC purposes. For example, an antenna with two RF ports, each of which excites one of the mutually orthogonal electromagnetic wave polarization, can be designed to suppress the leak (mutual coupling) between its ports. In fact, suppressing the RF leak between the antenna's ports can effectively be interpreted as antenna passive SIC from a full-duplex wireless transceiver perspective. A magneto-electric dipole (MED) antenna provides a suitable antenna structure type as a dual-port antenna for a full-duplex transceiver in network infrastructure -- such as a basestation. The MED suitability stems from possessing two dually polarized ports with broadband resonance and hemispherical radiation patterns \cite{Luk2006}. 


In spite of the MED's suitability for full-duplex network infrastructure applications due to its radiation pattern and broadband resonance, the isolation among its RF is the essential property. In the past, state-of-the-art MED focused on resonance bandwidth and fabrication methods; however, the RF port isolation was a secondary sub-optimized parameter as full-duplex scheme supportability was not a requirement. For example, in \cite{pcb1} and \cite{pcb2}, respectively, a printed circuit board (PCB) and low-temperature co-fired ceramic (LTCC)-based dual-polarized MED antennas presented for array's configurations but they have poor isolation levels and complex designs since they require via holes, multi-layers, etc. Other examples of dual-polarized MED antennas with metallic plates are presented in \cite{simpleFeeds, novelFeeds, widebandMED, novelMED}; however, the isolation level is still insufficient for full-duplex applications. Moreover, a MED for full-duplex applications was proposed in \cite{inBandFD}, but the ports isolation is limited to \SI{21}{dB} only.

This paper proposes a dual-polarized metal 3D printed MED with an orthogonal $\Gamma$-shape and a novel inverted $\Gamma$-shape probes. \emph{The antenna resonates over \SIrange[]{3}{4}{GHz} bandwidth with a stable radiation pattern around \SI{8}{dBi} maximum gain and more than \SI{50}{dB} ports' isolation level}. The structure of the paper is the following: the initial design of the MED is presented in Section~\ref{sec:initDesign}. Then, Section~\ref{sec:analysis} presents a parametric analysis of the MED design and investigation with respect to imperfect orthogonality between the MED's probes. The final design and experimental validation are presented in Sections~\ref{sec:finalDesign}. Finally, the paper concludes in Section~\ref{sec:conclusion}.

\section{MED Antenna Initial Design}
\label{sec:initDesign}

\subsection{Antenna Structure}
\label{subsec:antennaStructure}

\begin{figure*} [t]
\centering
\subfloat[3D model]{\includegraphics[width=0.5\columnwidth]{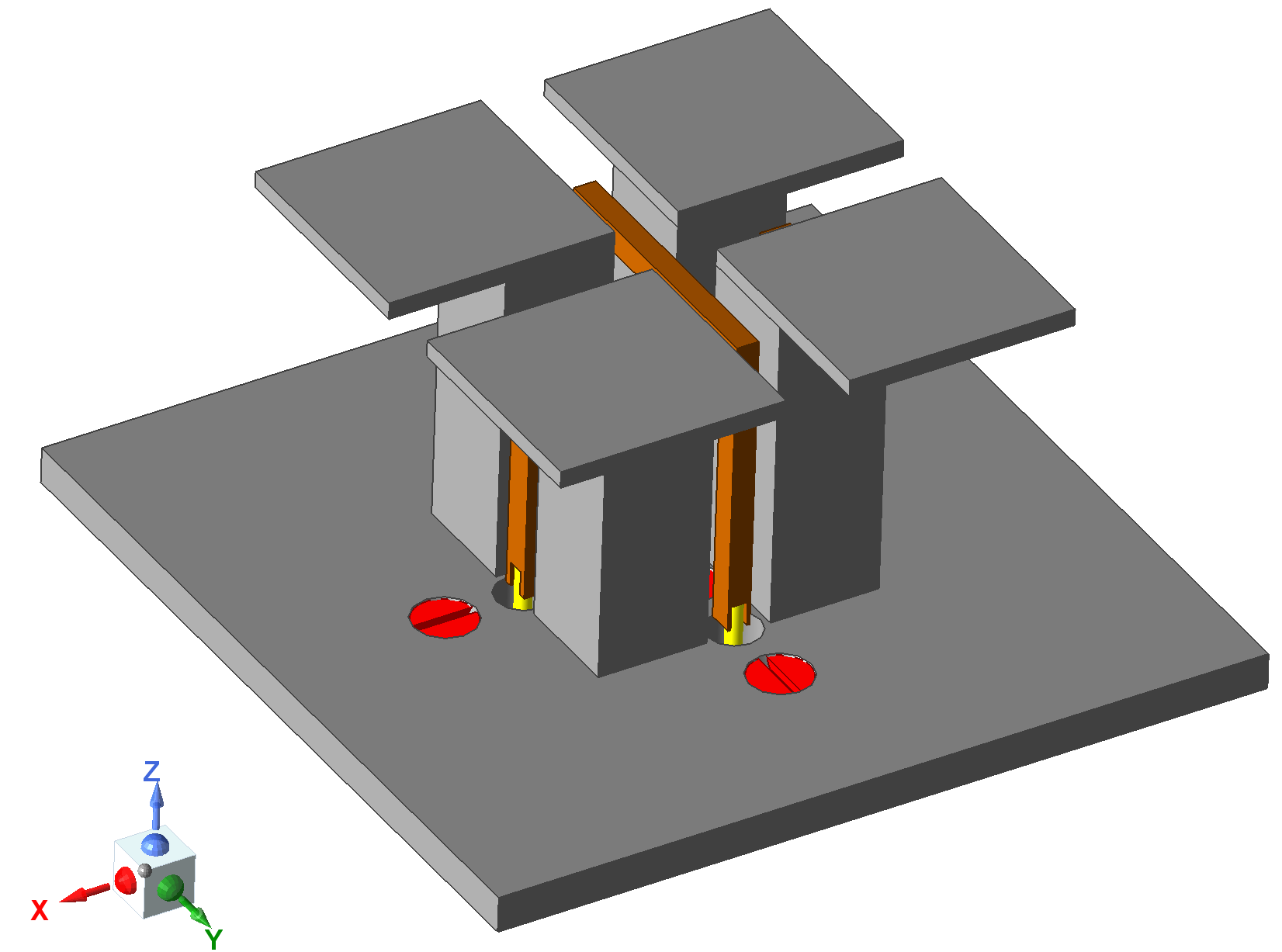}}
\hfil
\subfloat[Feed network]{\includegraphics[width=0.4\columnwidth]{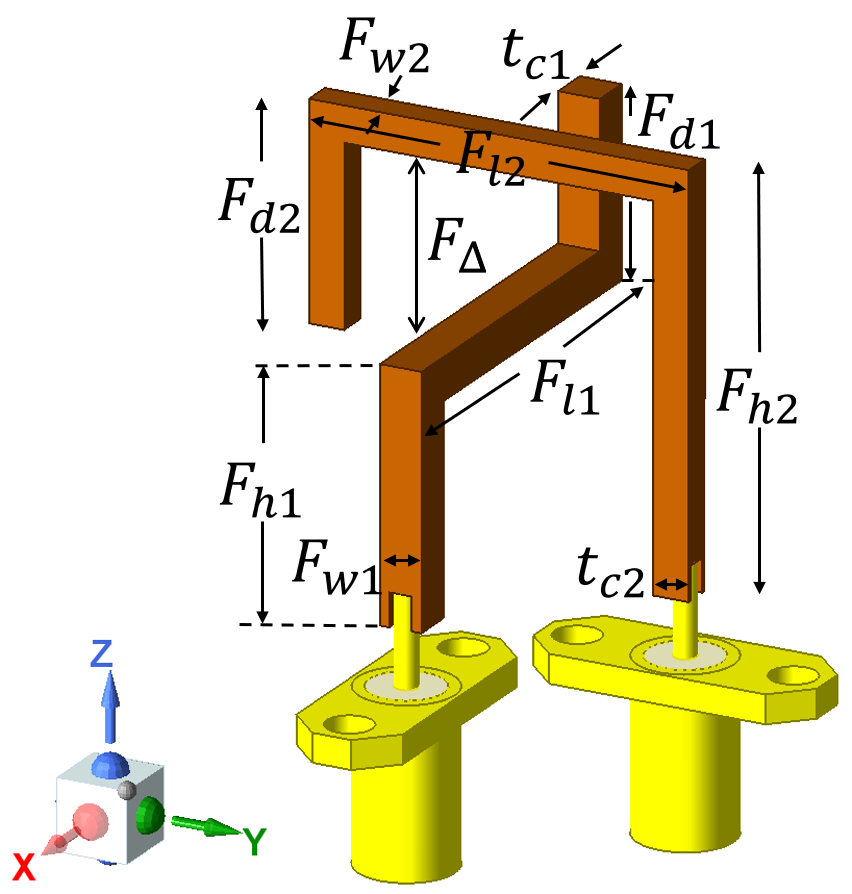}}
\\
\vspace{-0.3cm}
\subfloat[Top-view]{\includegraphics[width=0.4\columnwidth]{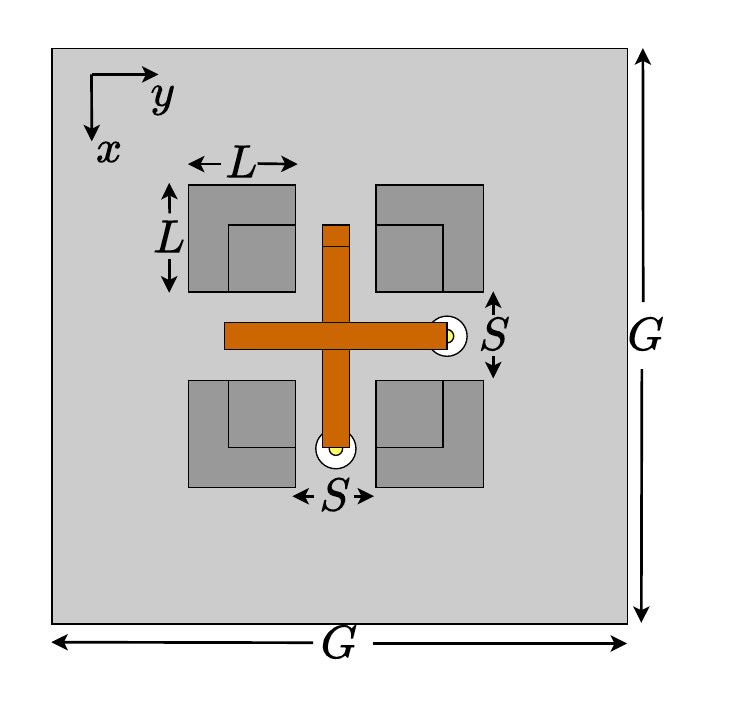}}
\hfil
\subfloat[Side-view (Port 1)]{\includegraphics[width=0.6\columnwidth]{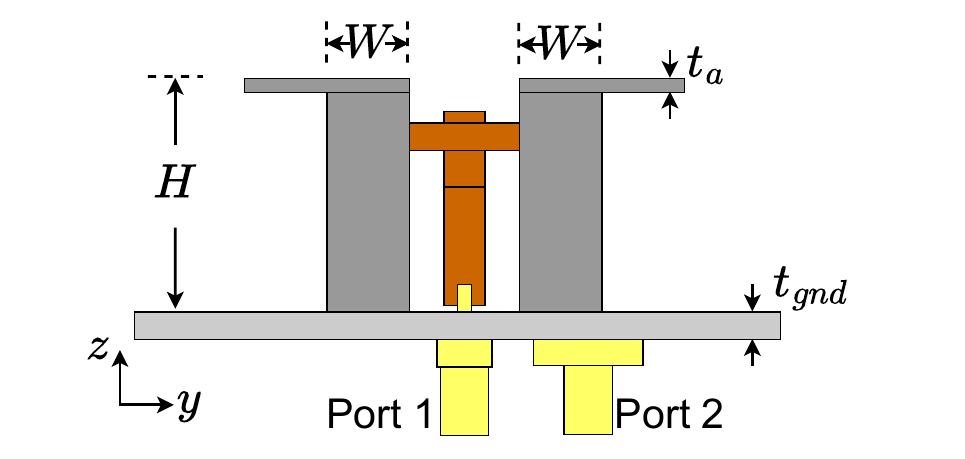}}
\hfil
\subfloat[Side-view (Port 2)]{\includegraphics[width=0.6\columnwidth]{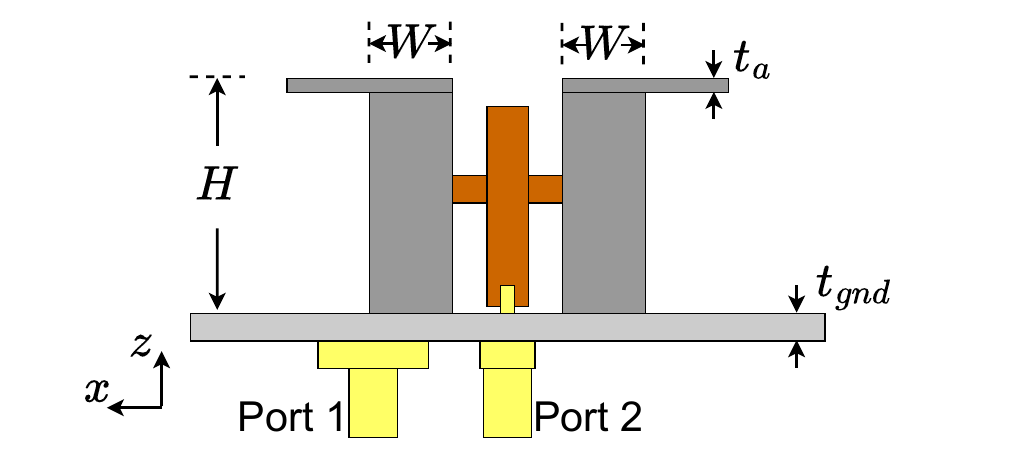}}
\caption{Dual-polarized MED structure.}
\vspace{-0.2cm}
\label{fig:initialMED}
\end{figure*}

\begin{table*}[bht]
\renewcommand{\arraystretch}{1.3}
\caption{MED Antenna Dimension Parameters and their values in \SI{}{mm}}
\label{tab:dimensions}
\centering
\begin{adjustbox}{width=\linewidth}
\begin{NiceTabular}{r | c c c c c c c c c | c c c c c c c c c } 
\CodeBefore
\rowcolor{gray!25}{1-2}
\Body
\toprule
\Block{2-1}{\textbf{Parameter}} & \Block{1-9}{\textbf{MED posts and plates}}  &&&&&&&&                          & \Block{1-9}{\textbf{MED feeds}} &&&&&&&&\\
                                \cmidrule(rl){2-10}                                                             \cmidrule(rl){11-19}
                                &$L$    &$S$     &$H$     &$G$ &$W$     &$t_a$  &$t_{c_1}$   &$t_{c_2}$  &$t_{gnd}$   &$F_{w_1}$   &$F_{w_2}$   &$F_{h_1}$   &$F_{h_2}$   &$F_{\ell_1}$   &$F_{\ell_2}$ & $F_{d_1}$ & $F_{d_2}$ &$F_\Delta$  \\
\textbf{Value / \SI{}{mm}}      &17.554 &4.971  &25.354 &60 &8.451  &1.3    &2.357      &1.997     &2.803       &2.357      &1.714      &13.971     &23.083     &21.429         &21.429       &10.5      &12        &7.114       \\
\bottomrule
\end{NiceTabular}
\vspace{-0.5cm}
\end{adjustbox}
\end{table*}

Fig. \ref{fig:initialMED} depicts the antenna structure. Four metallic plates and posts are used as a pair of half-wave electric dipoles and shorted quarter-wave patches (i.e. magnetic dipole), respectively. Two orthogonal probe feeds are used to excite the antenna. Both probes can be considered as three sections. The first section is the vertical line that transmits the signal from the input port to the horizontal section. This section can be seen as an air-filled strip line since the probe acts as an inner conductor, two posts act as a ground layer, and air acts as a substrate. Therefore, the feed width and air gap can be designed to obtain a specific characteristic impedance of the equivalent strip line.  The second section is the horizontal line which is the coupling element between the electric and magnetic dipoles. This section is very inductive, so it can make the antenna mismatched. Finally, the third section is the last vertical line that acts as an open-ended strip line. This section has a capacitive effect so that the inductive effect of the previous section can be compensated \cite{Luk2006}. To obtain a high isolation level, the probes must be as far from each other as possible. The previous designs from the literature have used two-orthogonal $\Gamma$-shaped probe feeds \cite{simpleFeeds}, \cite{widebandMED}. However, the open-ended section of the shorter probe may cause a geometrical constraint since its first section cannot be shortened to increase the gap between the feeds. Table \ref{tab:dimensions} summarizes all MED and its feed network dimensions.


\subsection{Electromagnetic Simulation Results}
\label{subsec:simREsults}
\begin{figure} [t]
\centering
\subfloat[SWR of both ports and isolation \label{fig:init_swr_iso}]{\includegraphics[width=0.5\columnwidth]{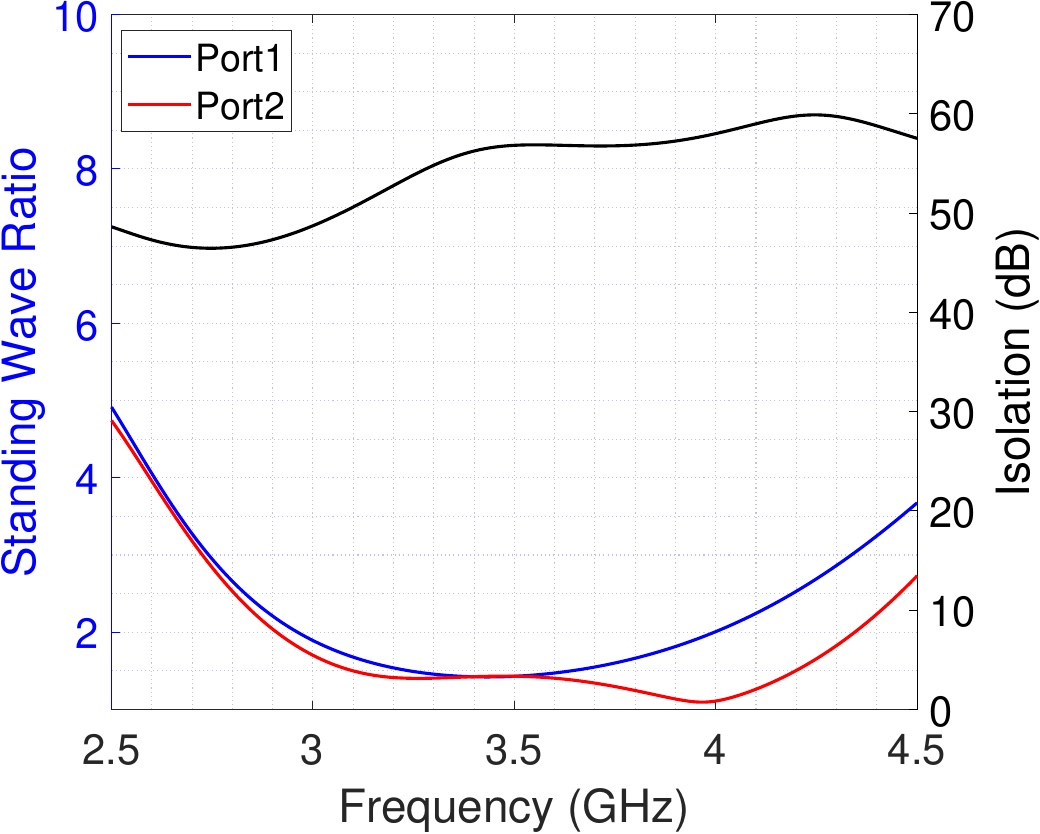}}
\hfil
\subfloat[Effects of $F_{h1}$ \label{fig:f_delta_effect}]{\includegraphics[width=0.5\columnwidth]{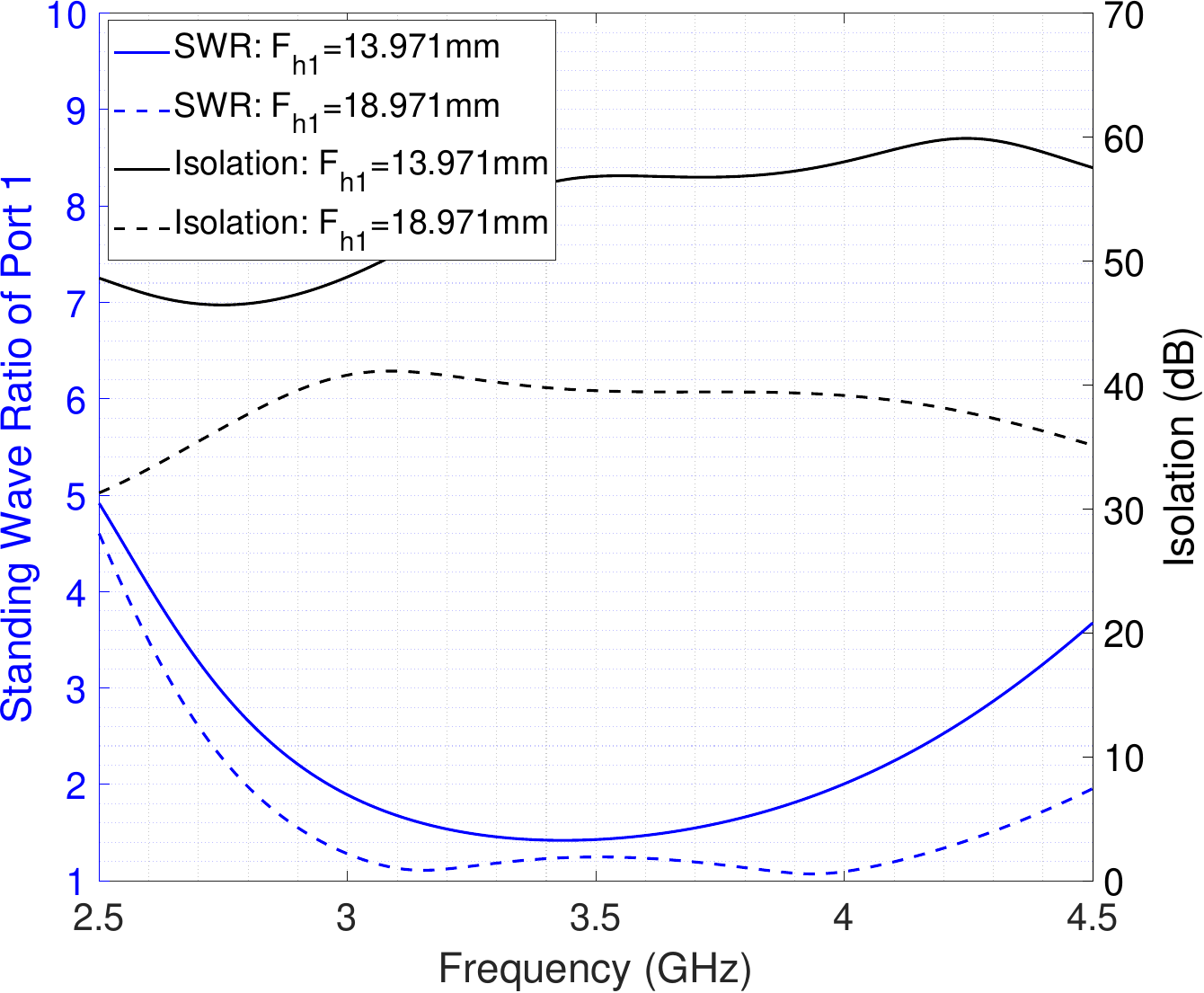}}
\caption{Full-wave simulation results of initial MED design.}
\vspace{-0.5cm}
\label{fig:init_med_fw_results}
\end{figure}

An electromagnetic (EM) full-wave simulation was done by using ANSYS\textsuperscript{\textregistered}~HFSS\textsuperscript{\texttrademark} (High-Frequency Structure Simulator) utilizing a finite element solver. Fig. \ref{fig:init_swr_iso} shows the simulation results. The left axis represents the standing wave ratio (SWR), and the blue and red lines represent the SWR values of port 1 and port 2, respectively. On the other hand, the black curve is the cross-polar isolation (XPI) between ports, and it is given in the right axis. It can be seen that SWR is less than 2 in the \SIrange[]{3}{4}{GHz} band, and isolation (XPI) is greater than \SI{50}{dB}. 

\section{Parametric Analysis}
\label{sec:analysis}

\subsection{First Probe Height}
\label{subsec:F_delta}

\begin{figure*} [t]
\centering
\subfloat[Forward tilt at Port 1 ($\alpha_1$)]{\includegraphics[width=0.5\columnwidth]{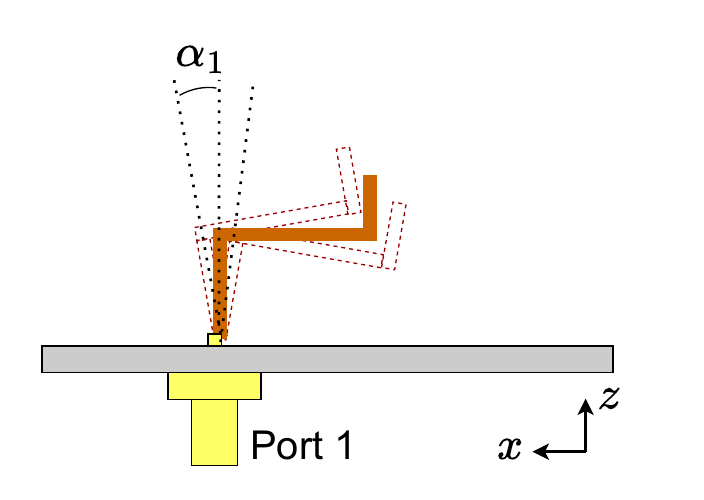}}
\hfil
\subfloat[Forward tilt at Port 2 ($\alpha_2$)]{\includegraphics[width=0.5\columnwidth]{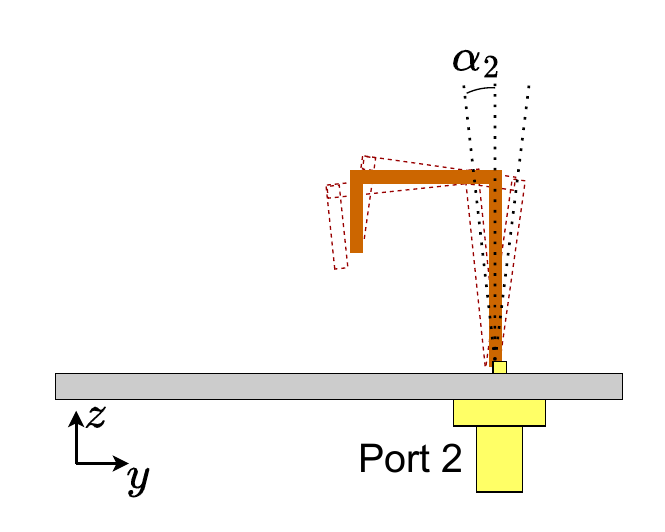}}
\hfil
\subfloat[Side tilt at Port 1 ($\beta_1$)]{\includegraphics[width=0.5\columnwidth]{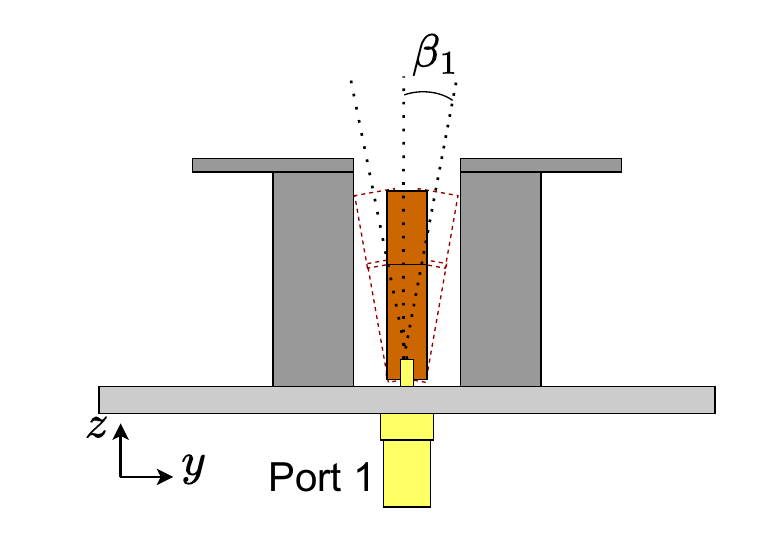}}
\hfil
\subfloat[Side tilt at Port 2 ($\beta_2$)]{\includegraphics[width=0.5\columnwidth]{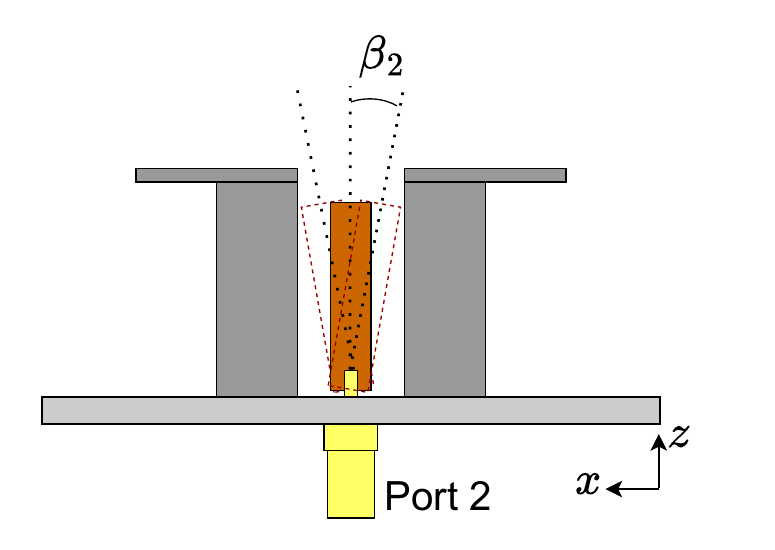}}
\caption{Illustration of MED's feed probes misalignment error angles.}
\vspace{-0.5cm}
\label{fig:errorAngles}
\end{figure*}

It is also seen from Fig. \ref{fig:init_swr_iso} that the impedance bandwidth of the first port is narrower than the second port. This is because the height of the first probe is smaller than the second probe. Therefore, the coupling point of the first probe is not at the optimum point. In other words, the coupling strength between the electric and magnetic dipoles is weak for the first probe due to its height. Fig. \ref{fig:f_delta_effect} illustrates the SWR and isolation for different heights of the first probe. It is clearly seen that when the height is increased, the coupling strength also increases, and the impedance bandwidth of the first probe gets wider. However, the isolation level is worsening since the two probes are getting closer (the gap is reduced). It ends up with a trade-off. Increasing the height of the first probe increases the coupling strength between the dipoles, providing a wider impedance bandwidth. However, the coupling between the probes is also increasing, and it reduces the isolation level. Since the impedance bandwidth of the first probe is large enough, the height was not changed to have a higher isolation level. 

\subsection{Probes Imperfect Orthogonality}
\label{subsec:orthogonality}
It is empirically predicted that the mutual orthogonality between the MED's probes may be affected by antenna assembly -- in technical terms, due to the soldering process of the probes to the RF coaxial connectors. Two error angles ($\alpha, \: \beta$) are defined as shown in Fig. \ref{fig:errorAngles} to analyze the imperfect probes orthogonality. The error angle $\alpha_i$ represents the tilting along the forward direction (towards the center of the antenna structure or reverse), while the angle $\beta_i$ represents the tilting along the side direction (towards metallic posts). The subscript $i = 1,2$ shows the probe ID. 

\begin{figure} [t]
\centering
\subfloat[First port ($\alpha_1=\pm 2^\circ$) \label{fig:forward_tilt_p1}]{\includegraphics[width=0.5\columnwidth]{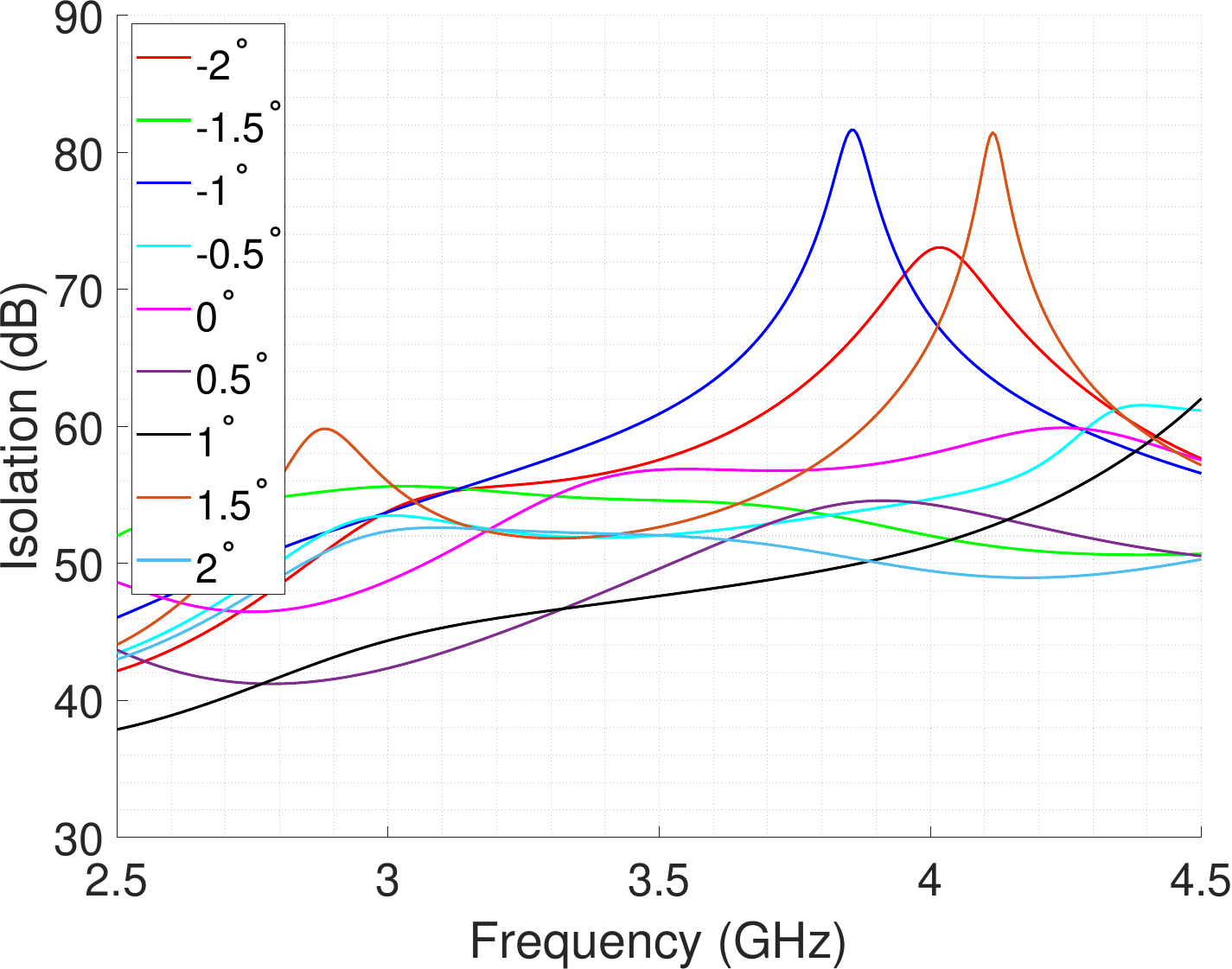}}
\hfil
\subfloat[Second port ($\alpha_2=\pm 2^\circ$) \label{fig:forward_tilt_p2}]{\includegraphics[width=0.5\columnwidth]{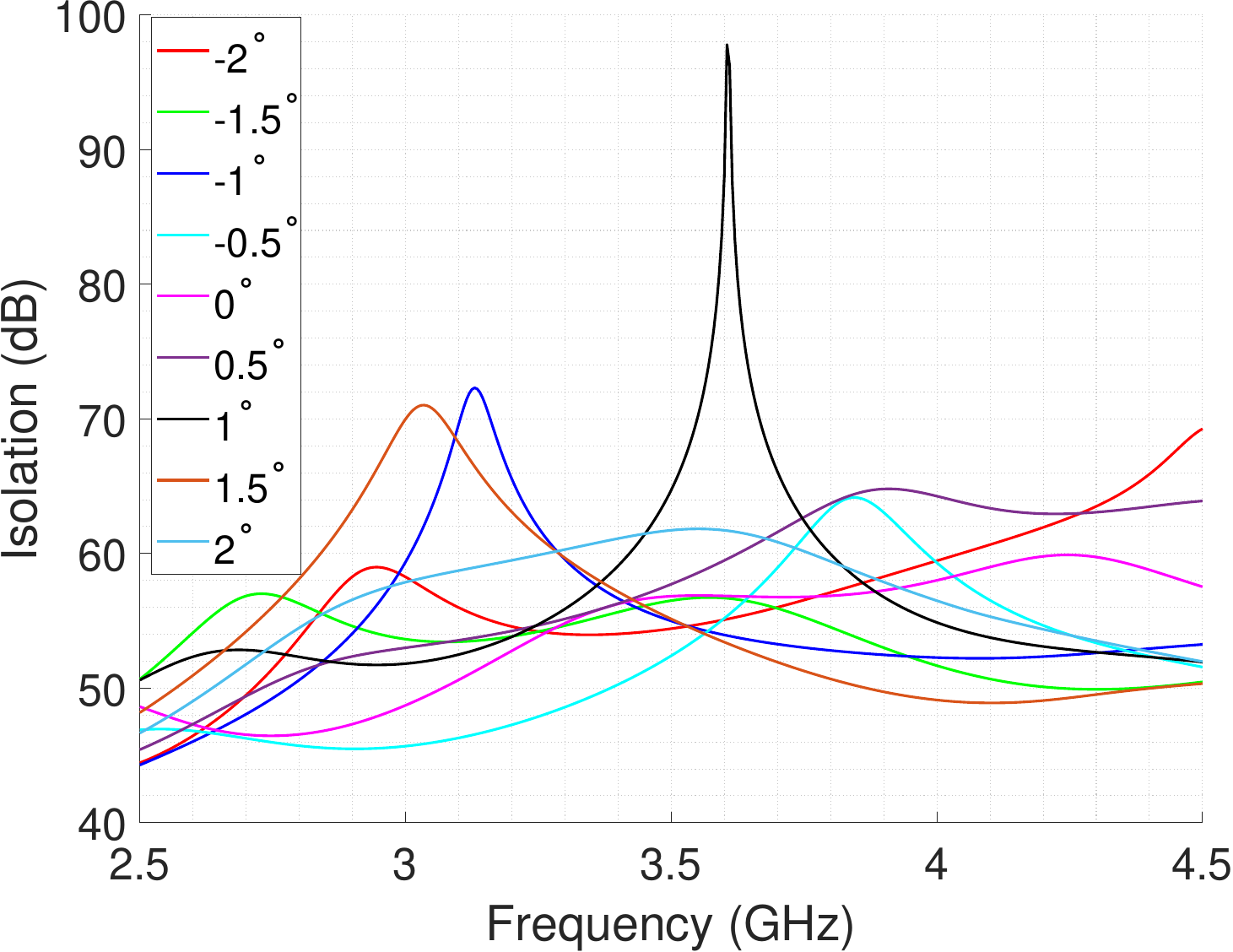}}
\caption{Effect of forward tilt error on isolation.}
\label{fig:forward_tilt}
\vspace{-0.35cm}
\end{figure}

\begin{figure} [t]
\centering
\subfloat[First port ($\beta_1=\pm 2^\circ$) ]{\includegraphics[width=0.5\columnwidth]{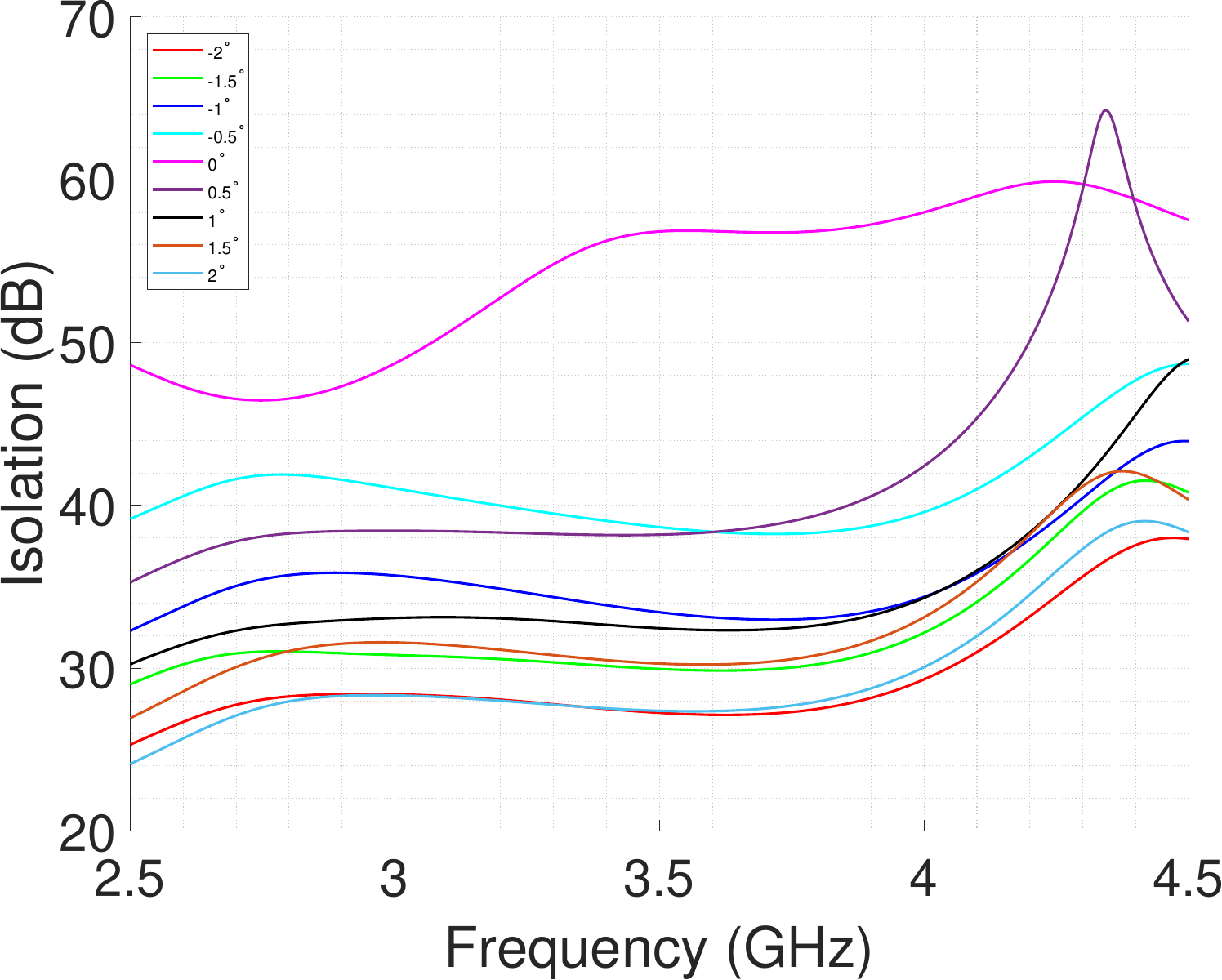}}
\hfil
\subfloat[Second port ($\beta_2=\pm 2^\circ$)]{\includegraphics[width=0.5\columnwidth]{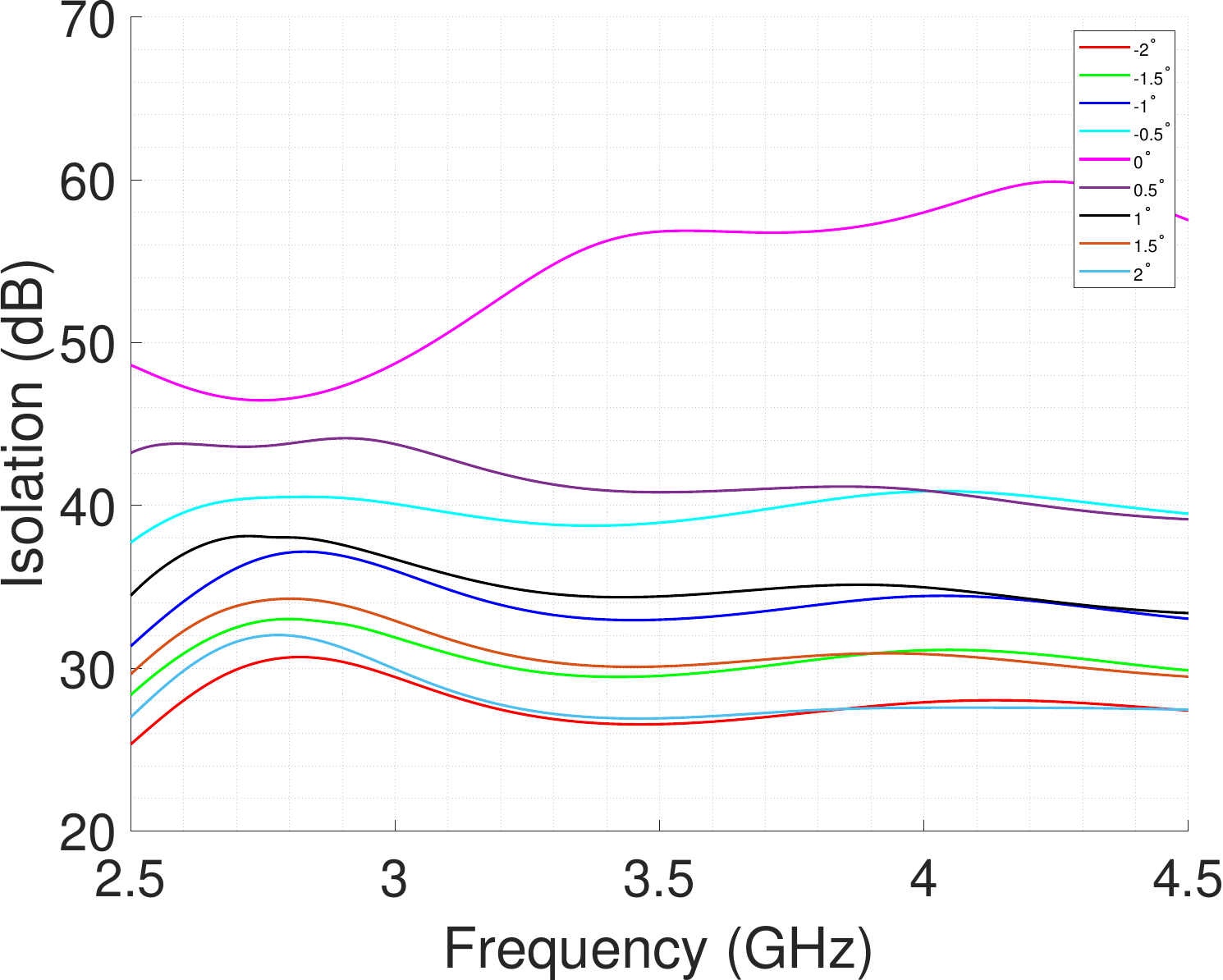}}
\caption{Effect of side tilt error on isolation.}
\label{fig:side_tilt}
\vspace{-0.3cm}
\end{figure}

Fig. \ref{fig:forward_tilt_p1} illustrates isolation for different forward tilting with $\alpha_1 = \pm 2^\circ$ tolerance on port 1 while all the other error angles set $0^\circ$. Although the isolation level is not 50 dB for all cases, it is still acceptable. A similar result has been obtained for the second port as shown in Fig. \ref{fig:forward_tilt_p2}.  A similar analysis has been done for the side tilting error, represented in Fig. \ref{fig:side_tilt}. Against the forward tilting, breaking the orthogonality by a side rotation reduces the isolation level dramatically.

 \begin{figure} [t]
\centering
\subfloat[Forward tilt ($\beta_{1,2}=0^\circ$) ]{\includegraphics[width=0.5\columnwidth]{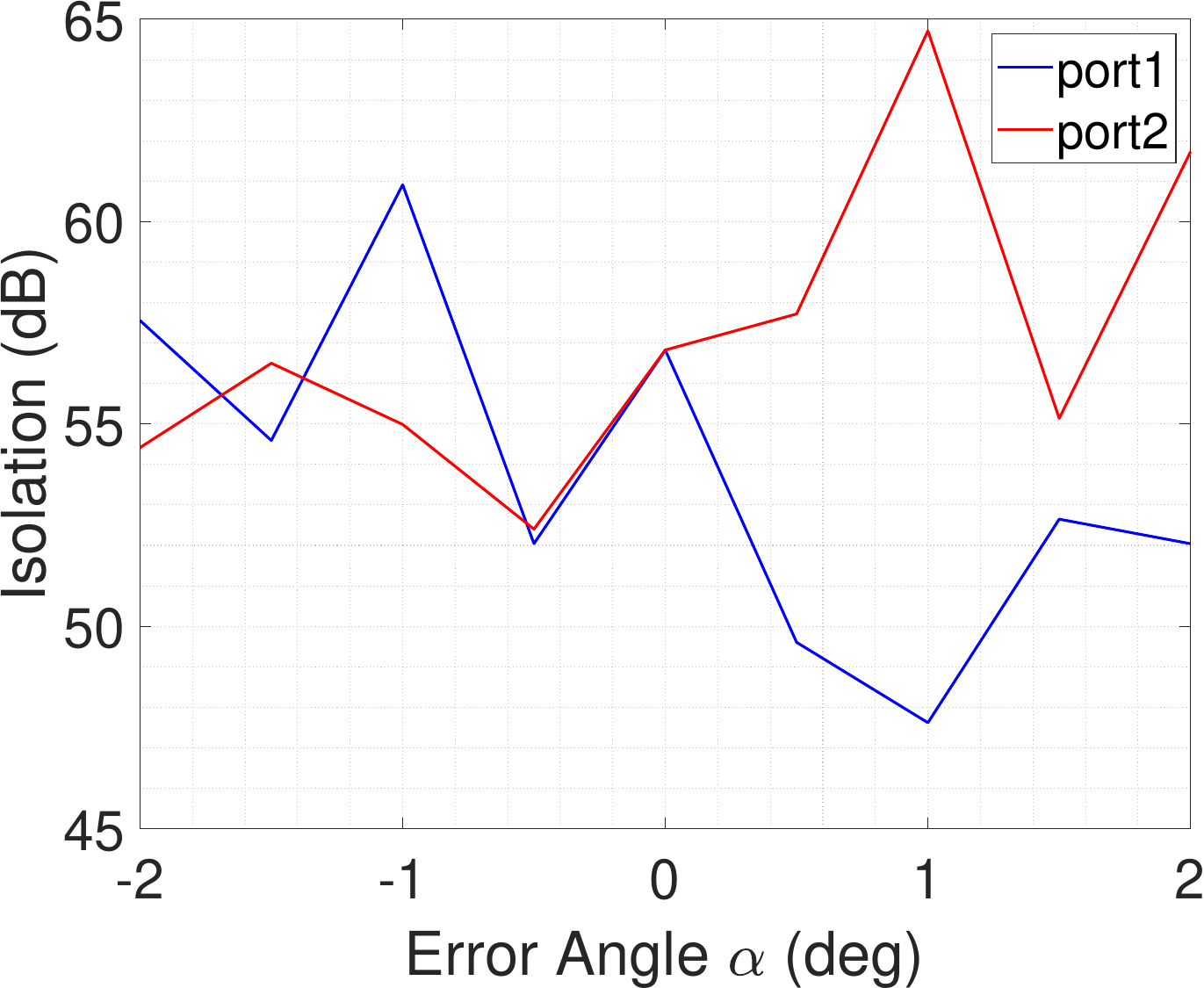}}
\hfil
\subfloat[Side tilt ($\alpha_{1,2}=0^\circ$)]{\includegraphics[width=0.5\columnwidth]{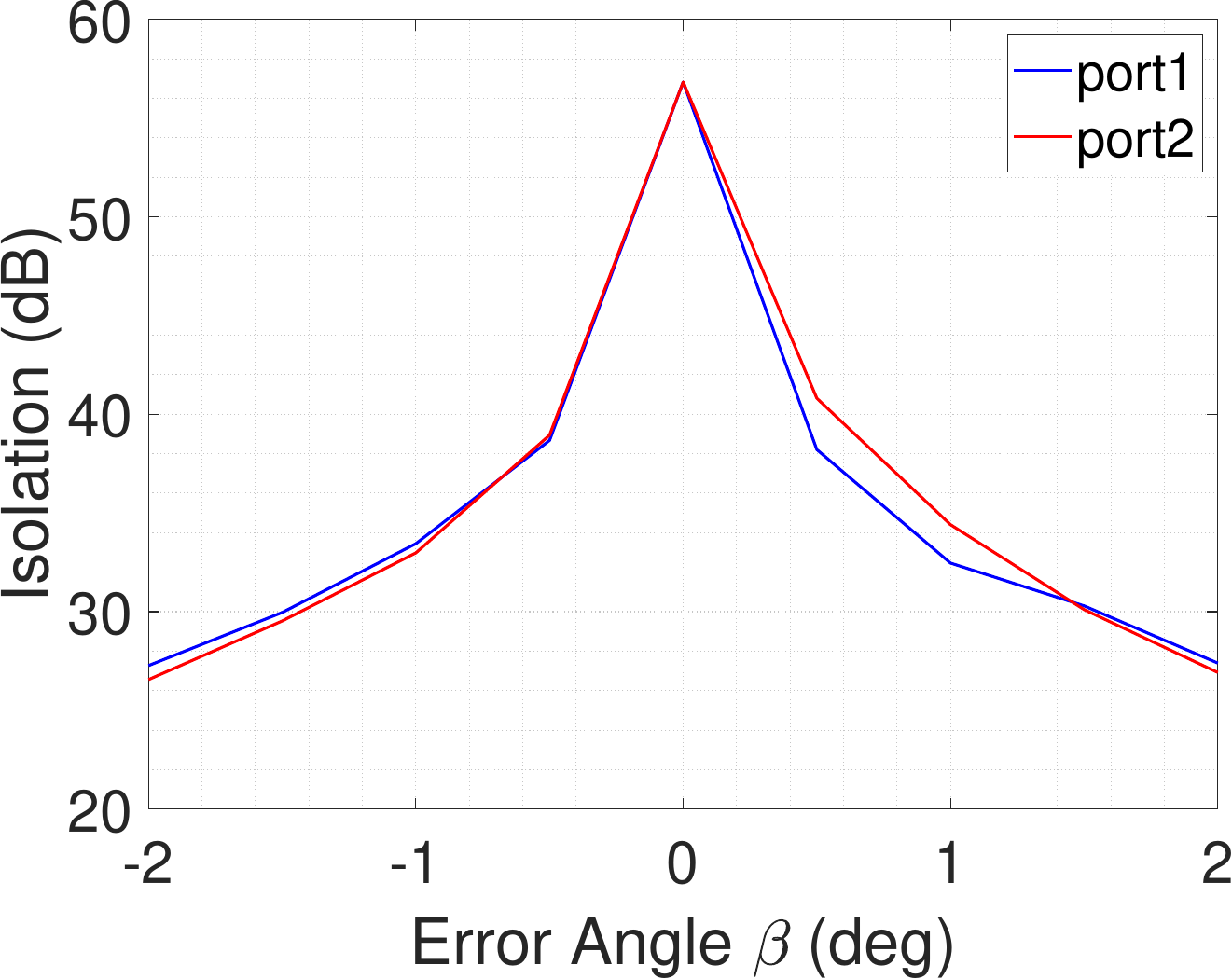}}
\caption{Isolation level versus error angles at 3.5 GHz.}
\label{fig:alpha__beta_center}
\vspace{-0.5cm}
\end{figure}

Finally, Fig. \ref{fig:alpha__beta_center} shows the isolation level at the center frequency for different values of $\alpha$ and $\beta$, respectively. It is seen that when the forward error angle is negative, the isolation level is better at some points for the first port, while it is better for the second port when the angle is positive. If we consider the global coordinate system, a negative forward tilting angle points to the center of the structure for the first port, on the contrary, a positive angle points to the center for the second port. Therefore, the isolation is better in these regions since the gap between the probes (i.e. $F_\Delta$) increases, and worse in the other areas since the gap decreases. On the other hand, it can be said that the effect of side tilting is almost identical for both ports and reciprocal for positive and negative angles.

\section{Final Design and Experimental Validation}
\label{sec:finalDesign}
\begin{figure} [t]
\centering
\subfloat[Side-view ]{\includegraphics[width=0.5\columnwidth]{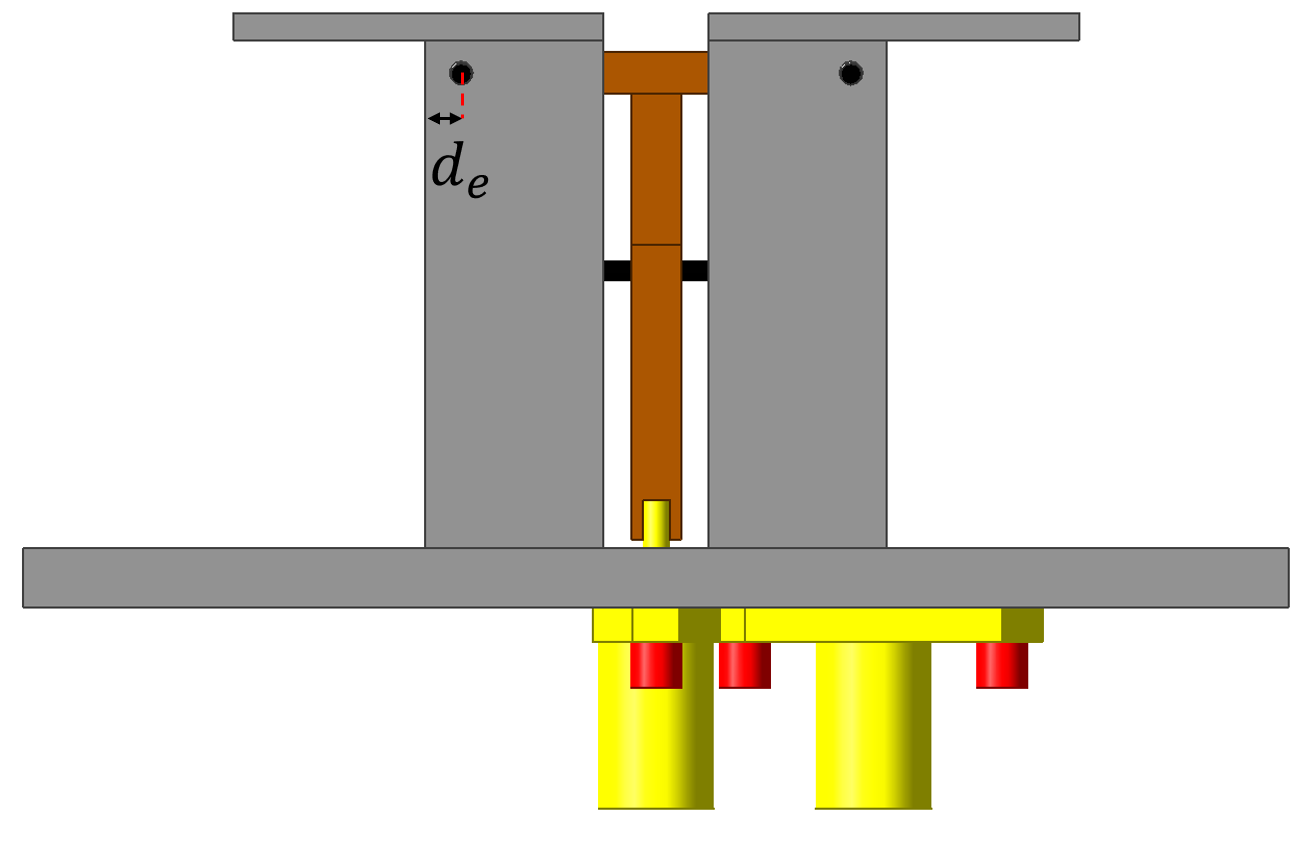}}
\hfil
\subfloat[Feed network]{\includegraphics[width=0.5\columnwidth]{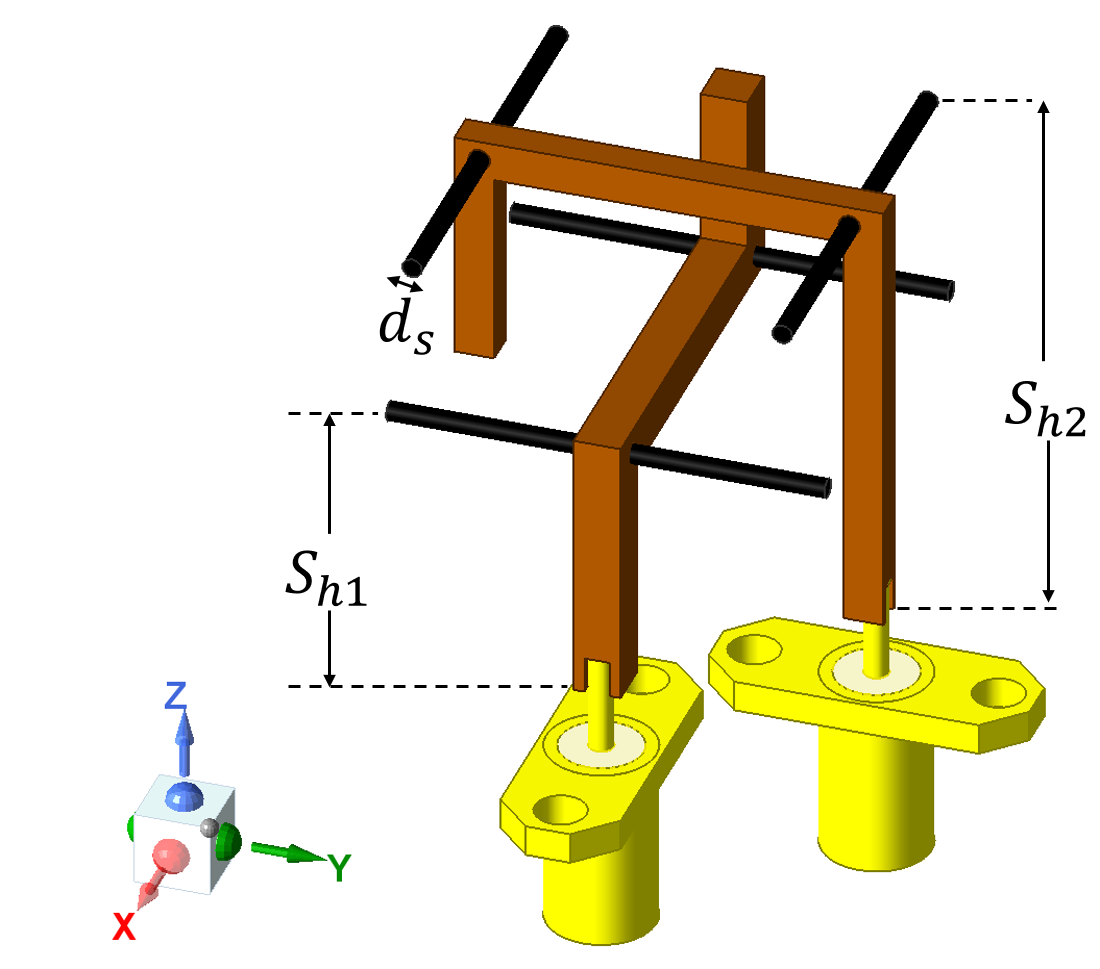}}
\caption{Improved MED design with dimensions (in mm): $t_{c1}=2.4347$, $S_{h1}=12.704$, $S_{h2}=22.034$, $d_s=1$, $d_e=1.7$ -- the rest of the parameters are the same.}
\label{fig:improved_Design}
\vspace{-0.5cm}
\end{figure}

After the parametric analysis, it was observed that the rotation error of the probes affected the performance. Due to this result, the initial design of dual-polarized MED has been improved to make the feeding network more stable mechanically. 4 pieces of plastic (Nylon PA12) sticks are placed through the metallic posts and probe feeds to hold the feeding network during and after the soldering process as shown in Fig. \ref{fig:improved_Design}. 

\begin{figure} [b]
\centering
\subfloat[Top-view]{\includegraphics[width=0.4\columnwidth]{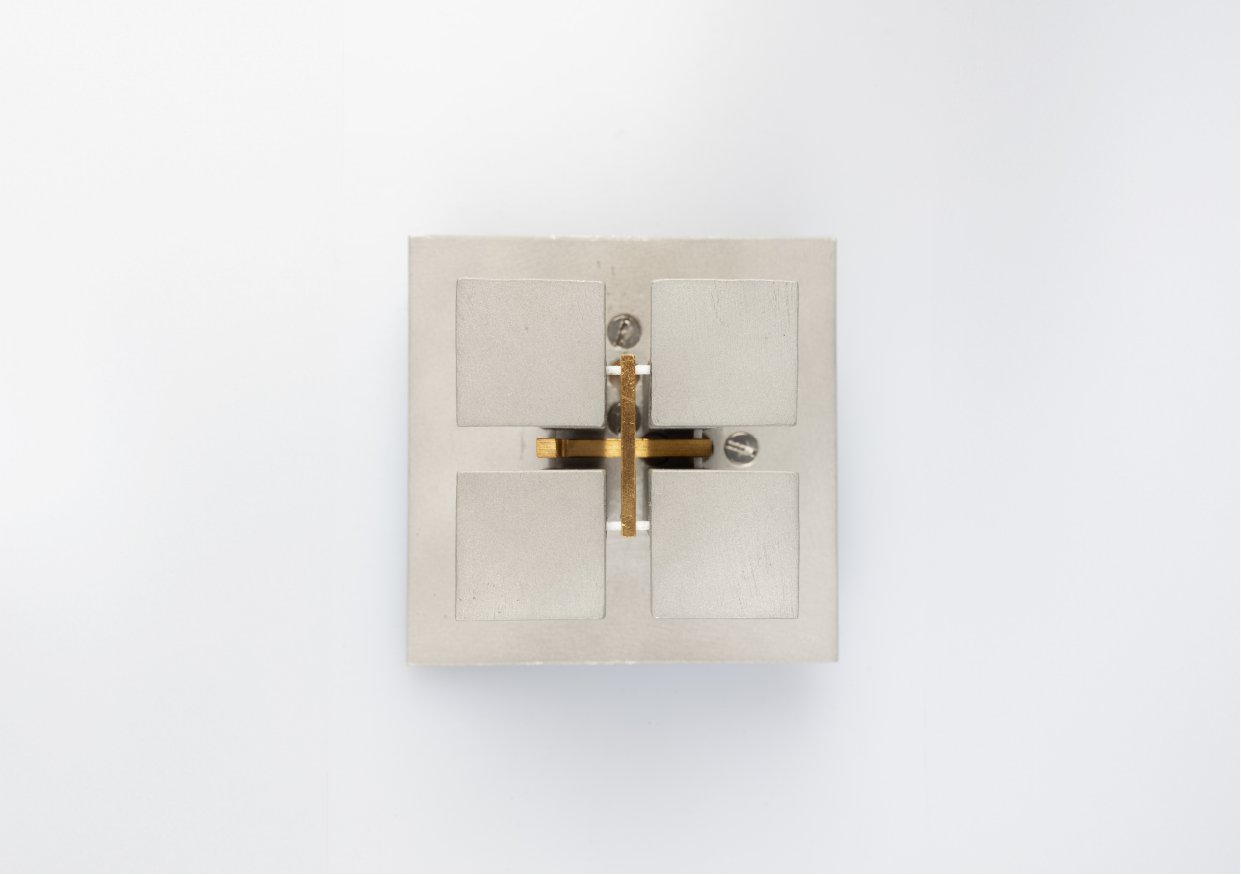}}
\hfil
\subfloat[Bottom-view]{\includegraphics[width=0.4\columnwidth]{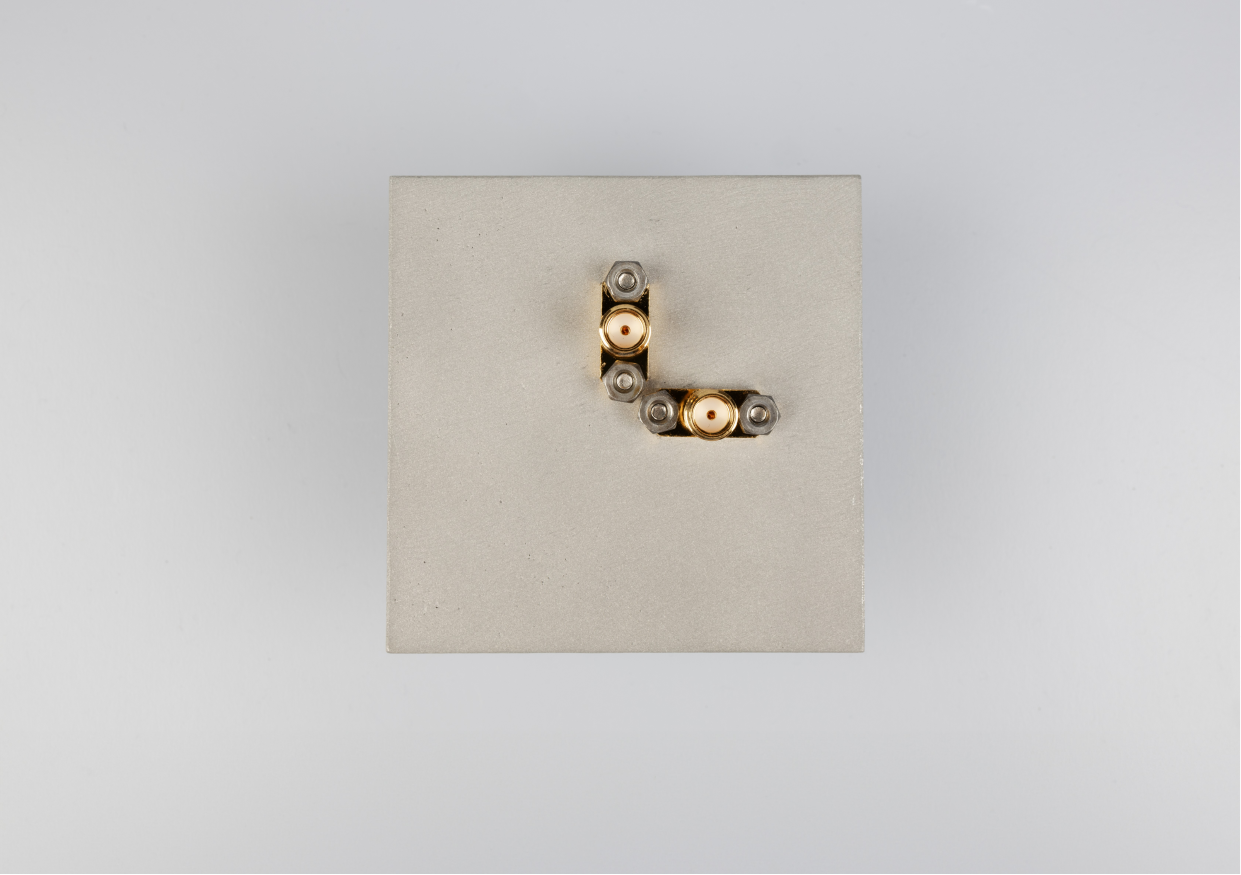}}
\\
\vspace{-0.1cm}
\subfloat[Perspective-view]{\includegraphics[width=0.5\columnwidth]{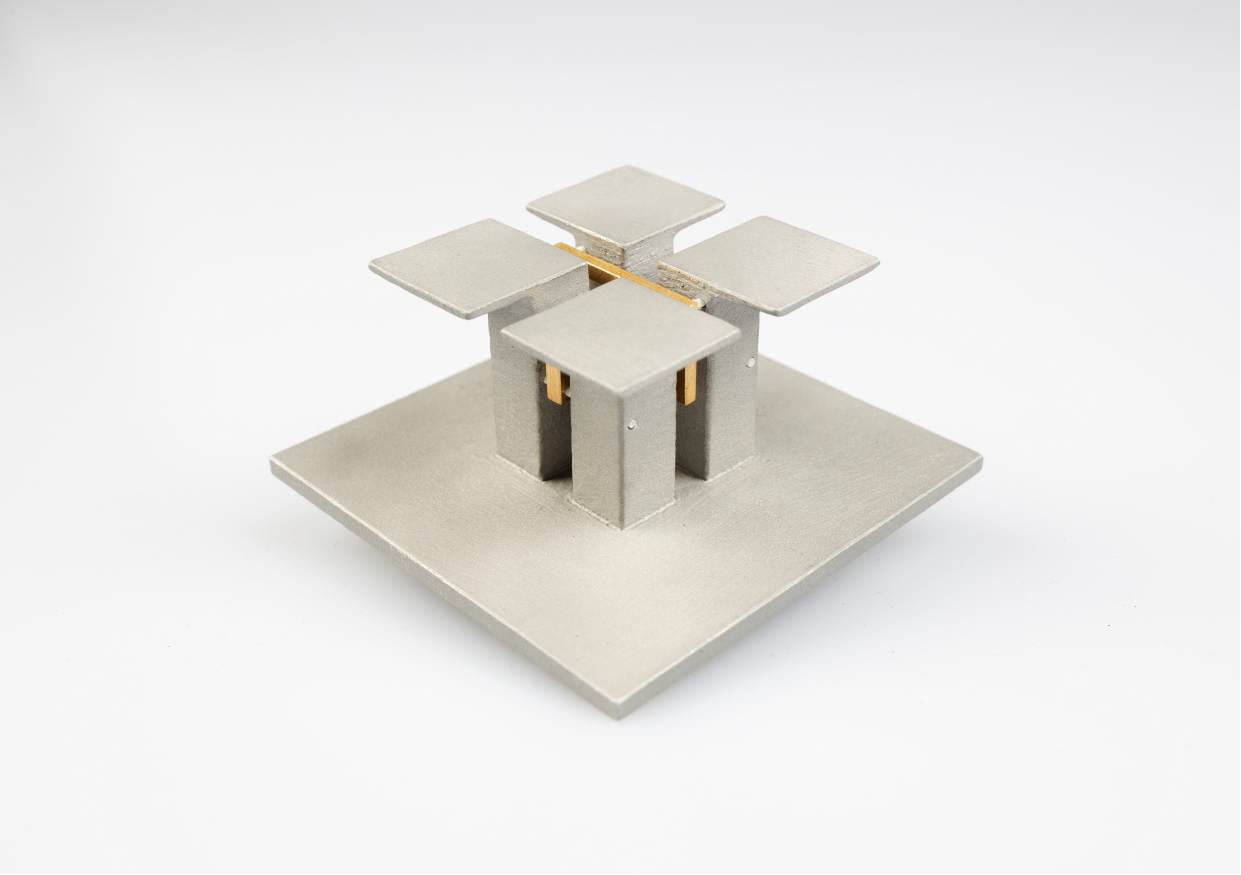}}
\caption{Photographs of the fabricated MED antenna.}
\label{fig:manufacturedMED}
\vspace{-0.5cm}
\end{figure}

The final design of the MED has been manufactured as given in Fig. \ref{fig:manufacturedMED}. The gray part of the antenna is 3D-printed Aluminum, white plastic sticks are 3D-printed Nylon (PA12), and probes are milled of brass. WR-SMA Panel Jack 2-Hole Flange Connector has been used to excite the antenna ports, and stainless steel DIN 963 countersunk screws with a slot have been preferred to mount the connectors.

\begin{figure} [t]
\centering
    \subfloat[SWR and Isolation \label{fig:meas_sim}]{\includegraphics[width=0.52\columnwidth]{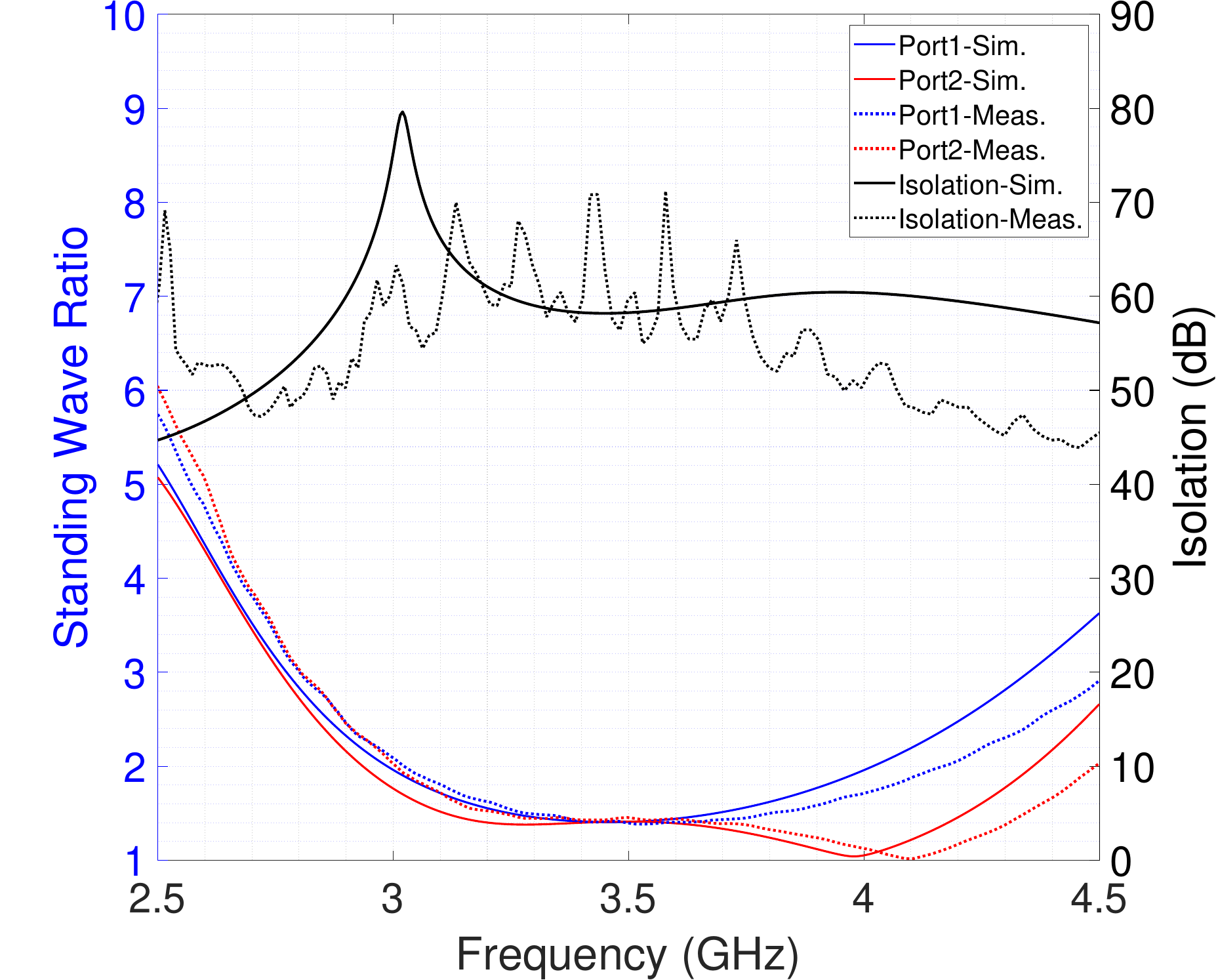}}
    \hfil
    \subfloat[Boresight gain \label{fig:gain_response}]{\includegraphics[width=0.48\columnwidth]{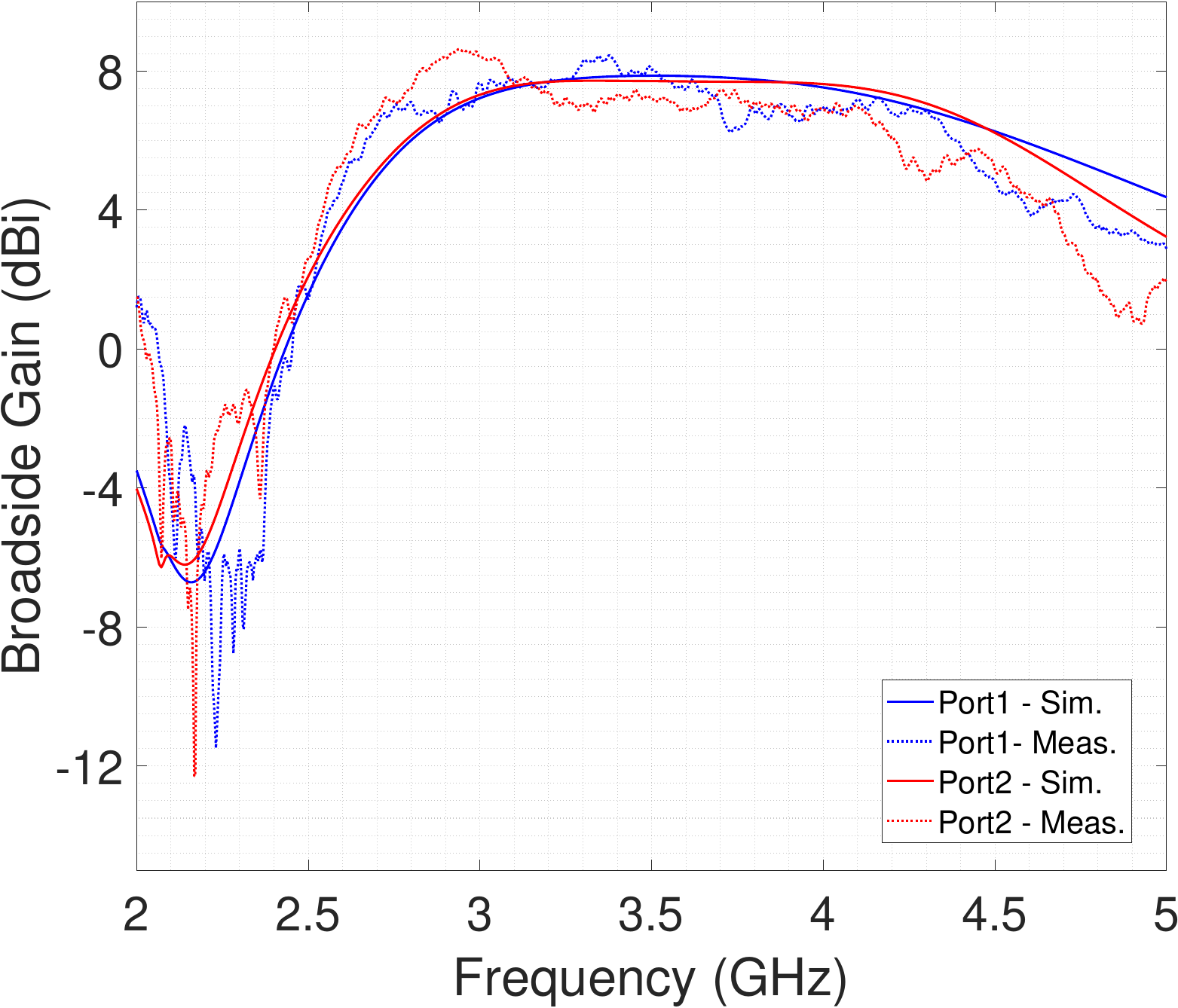}}
    \caption{Comparison between full-wave simulations and measurements.}
    \vspace{-0.3cm}
\label{fig:meas_sim_swr_gain}
\end{figure}

Fig. \ref{fig:meas_sim} illustrates the EM full-wave simulation results and measurements which were done by using the R\&S\textsuperscript{\textregistered} ZVA24 Vector Network Analyzer (VNA). It is observed from the graph that the SWR measurement and simulation results are in agreement. Although the isolation measurement is noisy due to the non-anechoic environment, the isolation level is more than \SI{50}{dB} within the \SIrange[]{3}{4}{GHz} range. It is also seen that SWR is less than 2 within the pass-band. Table \ref{tab:swr_sim_meas} summarizes the measurements and simulation results of SWR and isolation.

\begin{table} [t]
\renewcommand{\arraystretch}{1.3}
\caption{Comparison between simulated and measured SWR and isolation}
\label{tab:swr_sim_meas}
\centering
\begin{adjustbox}{width=\columnwidth}
\begin{NiceTabular}{ c c c c } 
\CodeBefore
\rowcolor{gray!50}{1-2}
\rowcolor{gray!10}{4}
\Body
        \toprule
                & \Block{1-2}{SWR $\leq$ 2 }  &                                       & \Block{2-1}{XPI\tabularnote{Cross-polar isolation (XPI) between port 1 and port 2.} $\geq$ 50 dB}   \\
                \cmidrule{2-3}
                & \Block{1-1}{Port 1 }  & \Block{1-1}{Port 2}   &     \\ 
    \midrule
    Simulated   & \SIrange{3}{4}{GHz}                        & \SIrange{2.9}{4.4}{GHz}                & \SI{>2.7}{GHz} \\
    Measured    & \SIrange{3}{4.2}{GHz}                     & \SIrange{3}{4.5}{GHz}                   &  \SIrange{2.9}{4.1}{GHz} \\
    \bottomrule
\end{NiceTabular}
\end{adjustbox}
\vspace{-0.5cm}
\end{table}

 Fig. \ref{fig:gain_response} shows both ports' boresight gain response with simulation and measurement data. It is seen that the designed MED antenna has a flat response within the desired frequency band. Since the second port has a wider impedance bandwidth, its gain response is also slightly wider than the first port. Moreover, simulation and measurements are almost in perfect agreement as the simulation shows that the boresight gain of the first and second ports are \SI{7.9}{dBi} and \SI{7.7}{dBi} at the center frequency, while the measurements are around \SI{8}{dBi} and \SI{7.2}{dBi}, respectively.

\begin{figure*}[t]
\centering
    \subfloat[Port 1 at 3 GHz]{\includegraphics[width=0.5\columnwidth, trim = 5cm 8.5cm 5cm 8cm, clip]{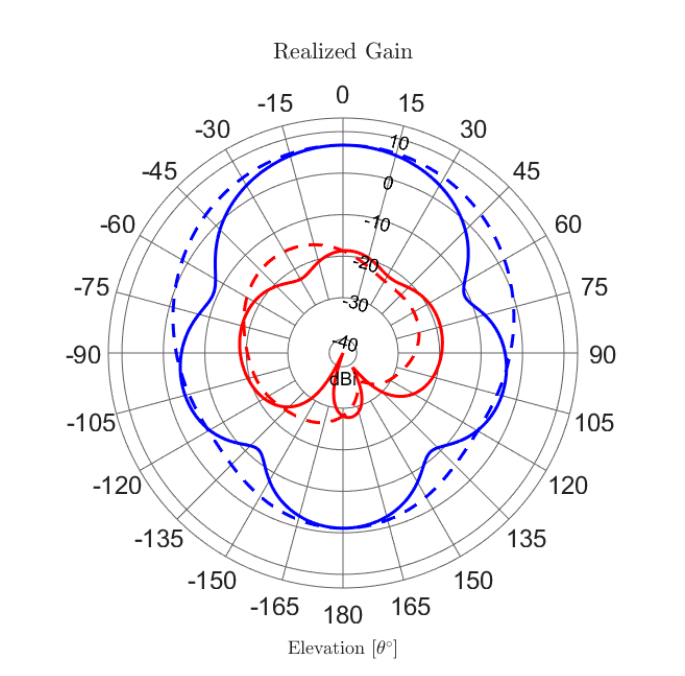}}
    \hfil
    \subfloat[Port 1 at 3.5 GHz]{\includegraphics[width=0.5\columnwidth, trim = 5cm 8.5cm 5cm 8cm, clip]{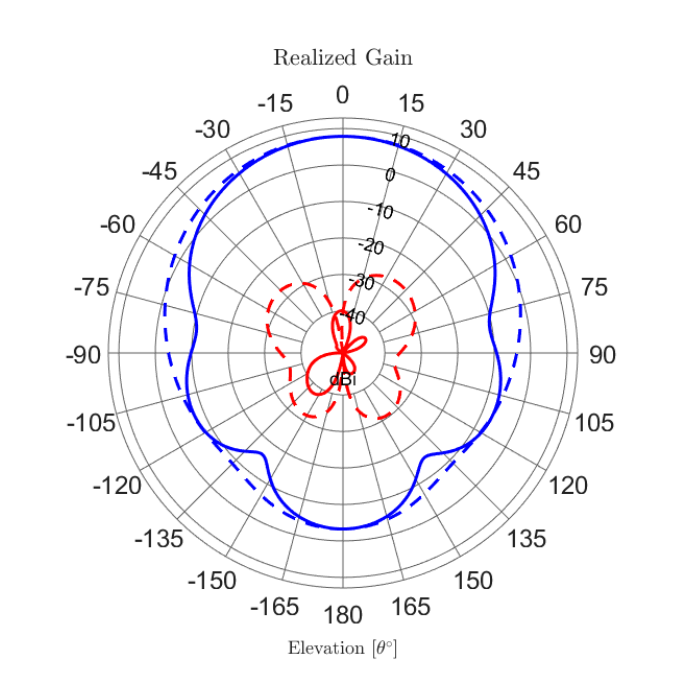}}
    \hfil
    \subfloat[Port 1 at 4 GHz]{\includegraphics[width=0.5\columnwidth, trim = 5cm 8.5cm 5cm 8cm, clip]{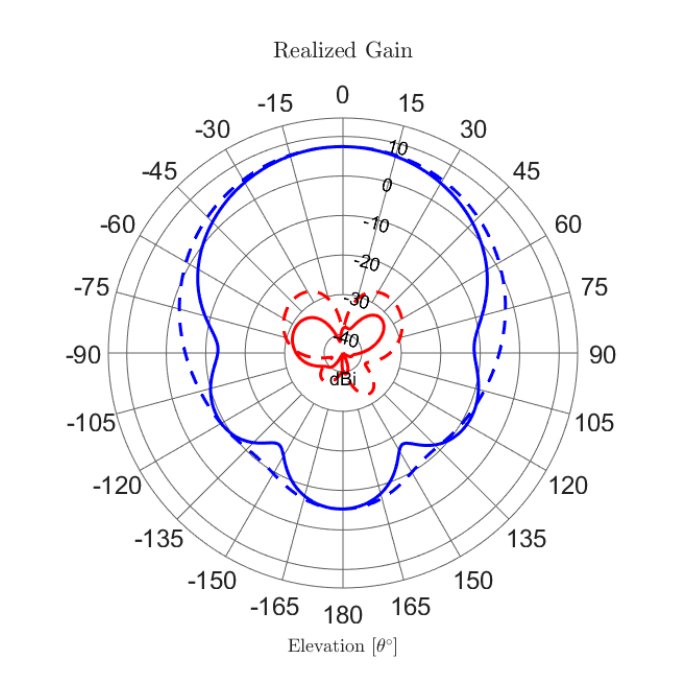}}
    \\
    \vspace{-0.5cm}
    \subfloat[Port 2 at 3 GHz]{\includegraphics[width=0.5\columnwidth, trim = 5cm 8.5cm 5cm 8cm, clip]{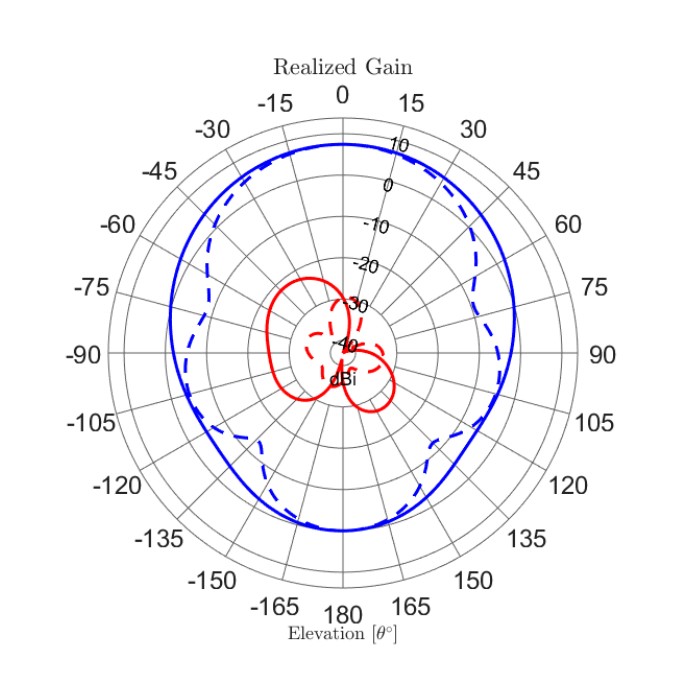}}
    \hfil
    \subfloat[Port 2 at 3.5 GHz]{\includegraphics[width=0.5\columnwidth, trim = 5cm 8.5cm 5cm 8cm, clip]{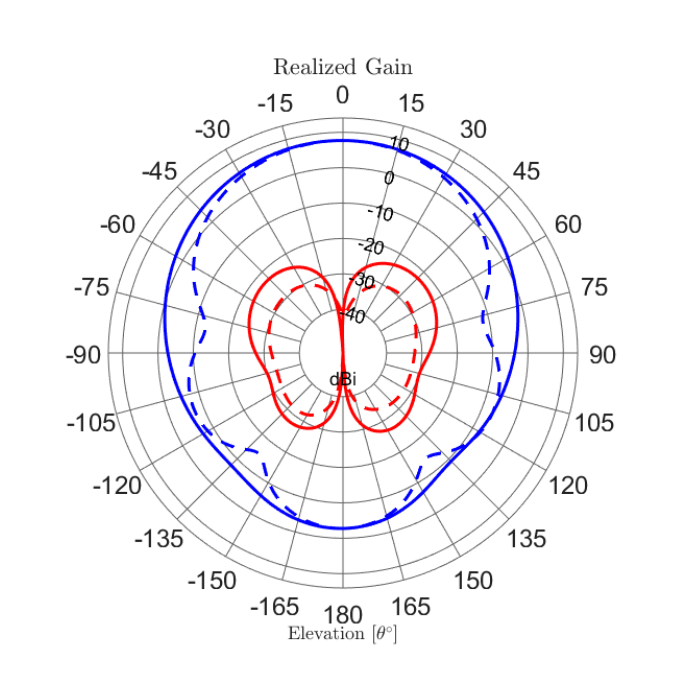}}
    \hfil
    \subfloat[Port 2 at 4 GHz]{\includegraphics[width=0.5\columnwidth, trim = 5cm 8.5cm 5cm 8cm, clip]{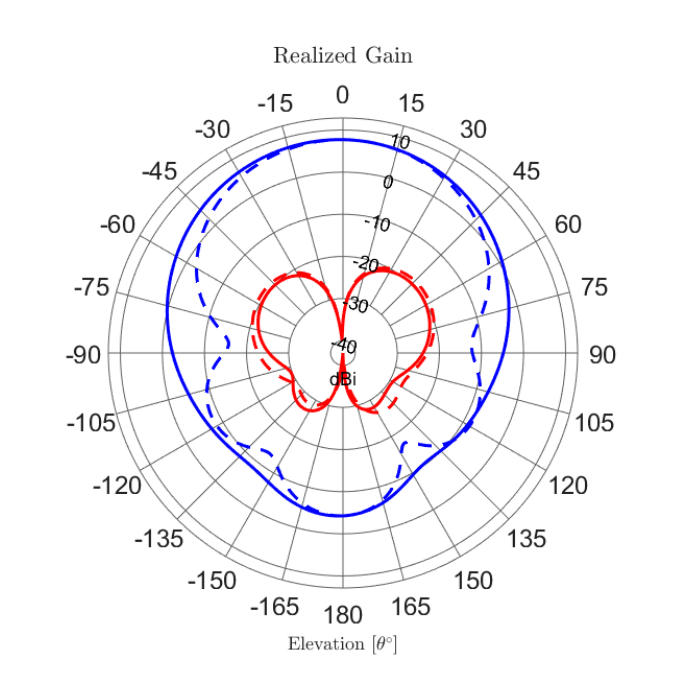}}
\caption{Full-wave simulation of radiation patterns at different frequency points. Solid lines: $\phi=0^\circ$ plane, dashed lines: $\phi=90^\circ$ plane -- blue: co-polarization, red: cross-polarization.}
\vspace{-0.5cm}
\label{fig:cross_x_gain}
\end{figure*}

\begin{table}[t]
\renewcommand{\arraystretch}{1.3}
\caption{Summary of simulated radiation patterns}
\label{tab:radiationPattern}
\centering
\begin{adjustbox}{width=\columnwidth}
\begin{NiceTabular}{  c c c c c c c  } 
\CodeBefore
\rowcolor{gray!50}{1-3}
\rowcolor{gray!10}{5}
\Body
    \toprule
    \Block{3-1}{\rotate Frequency}  & \Block{1-3}{Port 1} & & &\Block{1-3}{Port 2}  & &  \\
                        \cmidrule(rl){2-4} \cmidrule(rl){5-7} 
                        & \Block{1-2}{Half-power beamwidth} &               & \Block{2-1}{Realized \\gain}  &\Block{1-2}{Half-power beamwidth} &            & \Block{2-1}{Realized \\gain}  \\
                        \cmidrule(rl){2-3} \cmidrule(rl){5-6}
                        &  \Block{1-1}{$\phi=0^\circ$ \\plane} & \Block{1-1}{$\phi=90^\circ$ \\plane}     & &\Block{1-1}{$\phi=0^\circ$ \\plane}  &\Block{1-1}{$\phi=90^\circ$ \\plane}  &    \\
                \hline
                3 GHz       & \SI{62}{\degree}  & \SI{89}{\degree} & 7.2 dBi &  \SI{90}{\degree} & \SI{63}{\degree}  & 7.4 dBi \\
                3.5 GHz     & \SI{65}{\degree}  & \SI{83}{\degree} & 7.9 dBi &  \SI{86}{\degree} & \SI{64}{\degree}  & 7.7 dBi \\
                4 GHz       & \SI{67}{\degree}  & \SI{82}{\degree} & 7.4 dBi &  \SI{85}{\degree} & \SI{67}{\degree}  & 7.7 dBi \\
                \bottomrule
\end{NiceTabular}
\end{adjustbox}
\vspace{-0.3cm}
\end{table}

Fig. \ref{fig:cross_x_gain} represents the co-polarized and cross-polarized radiation patterns at three different frequency points for the excitation of ports individually. It is seen that the co-polarized realized gain at \SI{3.5}{GHz} is \SI{7.9}{dBi} while the cross-polarized gain is below \SI{-38}{dBi} and \SI{-28}{dBi} on $\phi = 0^\circ$ and $\phi = 90^\circ$ planes, respectively. Simulation results of the radiation patterns are listed in Table \ref{tab:radiationPattern}. Since the antenna is dual-polarized (vertical and horizontal), the radiation patterns of the ports are switched between the two cut-planes of $\phi=0^\circ$ and $\phi=90^\circ$. Moreover, the radiation pattern is stable within the desired bandwidth.

As a final point, Table \ref{tab:comparison_works} compares the proposed MED antenna to the state-of-the-art MED antennas. It is clearly seen that the proposed antenna has the highest isolation (XPI) level and the most stable gain within the required impedance bandwidth. Therefore, it is highly suitable for full-duplex systems. Although the proposed antenna has the narrowest impedance bandwidth, it has the advantage of easy manufacturing, which mainly depends on metal 3D printing.

\begin{table}[t]
\renewcommand{\arraystretch}{1.3}
\caption{Comparison between the proposed MED antenna and state-of-the-art ones}
\label{tab:comparison_works}
\centering
\begin{adjustbox}{width=\columnwidth}
\begin{NiceTabular}{  c c c c c  } 
    \CodeBefore
    \rowcolor{gray!50}{1}
    \rowcolor{gray!10}{2}
    \rowcolor{gray!10}{4}
    \rowcolor{gray!20}{6}
    \Body
    \toprule
    \RowStyle{\bfseries}
    Reference           & Fabrication       & Bandwidth                                             & Isolation         & Gain  \\
    \midrule
    \cite{pcb1}         & PCB               & \SIrange{24}{40}{GHz}\tabularnote{With respect to SWR $\leq$ 1.5.}     & \SI{>17.8}{dB}    & \SIrange{7}{9.4}{dBi}  \\
    \cite{widebandMED}  & Metallic plates   & \SIrange{1.68}{2.71}{GHz}\tabularnote{With respect to SWR $\leq$ 1.5.} & \SI{>30}{dB}      & \SIrange{7.6}{9.4}{dBi} \\
    \cite{novelMED}     & Metallic plates   & \SIrange{1.62}{2.87}{GHz}\tabularnote{With respect to SWR $\leq$ 1.5.} & \SI{>30}{dB}      & \SIrange{5.8}{8.9}{dBi} \\
    \cite{inBandFD}     & Metallic plates   & \SIrange{1.9}{2.75}{GHz}\tabularnote{With respect to SWR $\leq$ 1.5.}  & \SI{>20.9}{dB}    & \SI{>5.2}{dBi} \\
    \emph{This work}    & Metal 3D printing & \SIrange{3}{4}{GHz}\tabularnote{With respect to SWR $\leq$ 2.}         & \SI{>50}{dB}      & \SIrange{7.2}{7.9}{dBi} \\
    \bottomrule
\end{NiceTabular}
\end{adjustbox}
\vspace{-0.6cm}
\end{table}

\section{Conclusion}
\label{sec:conclusion}
A novel metal dual-port dual-polarized MED antenna is presented with orthogonal combined $\Gamma$ and inverted-$\Gamma$ shape probes. The antenna has a matched impedance bandwidth from \SIrange[]{3}{4}{GHz} with SWR $\leq$ 2, and an excellent XPI isolation performance that exceeds \SI{50}{dB} over the entire impedance bandwidth. Moreover, the antenna has a stable \SI{\pm 0.5}{dB} radiation pattern within its resonance frequency band and has a maximum gain of \SI{7.9}{dBi}. Therefore, the proposed antenna is a strong eligible candidate for full-duplex wireless communication to meet the greedy requirement of RF self-interference cancellation.

\section*{Acknowledgment}
The Federal Ministry of Education and Research of Germany (BMBF) supported this work through the project “6G-ICAS4Mobility”; the project has an identification number: 16KISK235.

\balance

\bibliographystyle{IEEEtran}
\bibliography{bibliography.bib}

%
‌%
‌%
‌%
‌%
%
%
%
%
‌%
\end{document}